\documentclass[twocolumn]{aastex631}
\usepackage{hyperref}
\usepackage{graphicx}	
\usepackage{amsmath}	
\usepackage{amssymb}	
\usepackage{newtxtext,newtxmath}
\usepackage{booktabs}
\usepackage{ragged2e}
\usepackage[mathlines]{lineno}

\newcommand{\revise}[1]{\textcolor{black}{#1}}
\newcommand{\reviseagain}[1]{\textcolor{black}{#1}}

\newcommand{\astrid}{\textsc{ASTRID}}
\newcommand{\Msun}{$ \mathrm{M}_{\odot}$}
\begin{document}

\title{Central Cluster Galaxies: A Hotspot for Detectable Gravitational Waves from Black Hole Mergers 
}


\author[0000-0002-8828-8461]{Yihao Zhou}
\affiliation{McWilliams Center for Cosmology, Department of Physics, Carnegie Mellon University, Pittsburgh, PA 15213, USA}

\author{Tiziana Di Matteo}
\affiliation{McWilliams Center for Cosmology, Department of Physics, Carnegie Mellon University, Pittsburgh, PA 15213, USA}

\author{Nianyi Chen}
\affiliation{School of Natural Sciences, Institute for Advanced Study, Princeton, NJ 08540, USA}
\affiliation{McWilliams Center for Cosmology, Department of Physics, Carnegie Mellon University, Pittsburgh, PA 15213, USA}

\author {Luke Zoltan Kelley}
\affiliation{Department of Astronomy, University of California, Berkeley, Berkeley, CA 94720, USA}

\author {Laura Blecha}
\affiliation{Physics Department, University of Florida, Gainesville, FL 32611, USA}

\author{Yueying Ni}
\affiliation{Center for Astrophysics $\vert$ Harvard \& Smithsonian, Cambridge, MA 02138, US}

\author{Simeon Bird}
\affiliation{Department of Physics \& Astronomy, University of California, Riverside, 900 University Ave., Riverside, CA 92521, USA}

\author{Yanhui Yang}
\affiliation{Department of Physics \& Astronomy, University of California, Riverside, 900 University Ave., Riverside, CA 92521, USA}

\author{Rupert Croft}
\affiliation{McWilliams Center for Cosmology, Department of Physics, Carnegie Mellon University, Pittsburgh, PA 15213, USA}



\begin{abstract}
After Pulsar Timing Arrays (PTAs) have announced the evidence for a low-frequency gravitational wave background (GWB), the continuous waves (CWs) are the next anticipated gravitational wave (GW) signals. 
In this work, we model CW sources detectable by PTAs based on the massive black hole (MBH) merger population in the ASTRID cosmological simulation.
We evolve MBH binaries, simulate their GW emissions, and calculate their detection probability (DP) for PTAs. 
The most detectable CW sources are produced by MBH mergers with masses $M_{\mathrm{BH}} > 10^{10}\, \mathrm{M}_{\odot}$ in the lowest frequency bins with $f<10$ nHz.
Remarkably, these mergers occur within massive galaxies with the stellar mass $M_{*}>10^{12}\, \mathrm{M}_{\odot}$ located at the center of galaxy clusters. 
Particularly striking in ASTRID is a triple merger event, wherein two consecutive mergers occur within 500 Myr interval in the same cluster core, generating high-DP CW signals at $\sim$ 2nHz and $\sim$ 10nHz. 
We also investigate the electromagnetic (EM) signatures associated with these events: either single or dual active galactic nuclei (AGN) in the massive host galaxies that are undergoing star formation. 
This research provides new insights into the low-frequency GW sky and informs future multi-messenger searches for PTA CW sources.



\end{abstract}



\section{Introduction} \label{sec:intro}
Massive black holes (MBHs) at the heart of massive galaxies are predicted to form binaries as a result of galaxy merging. Two MBHs sink to the center of the remnant galaxy via dynamical friction, and then their orbit continues to decay, reaching  $\sim 10^{-2}$ pc scales through 
a combination of stellar scattering and circumbinary disk torques. 
Below these scales,  MBH binary hardening is dominated by the emission of gravitational waves (GWs). Such GW signals are the main target for Pulsar Timing Arrays (PTAs) and the Laser Interferometer Space Antenna (LISA), which both open a new window other than traditional electromagnetic (EM) information to understand the population of MBHs. 

Multiple PTA collaborations have recently reported the first detection of GWs in the nanohertz frequency band (NANOGrav \citep{Agazie2023_Nanograv_GWB}; CPTA \citep{ Xu2023_CPTA}; PPTA \citep{Reardon2023_PPTA}; EPTA+InPTA \citep{EPTACollaboration2023}). Despite their different modeling choices, their results are consistent with each other, agreeing within 1$\sigma$ \citep{Agazie2024_PTAcompare}. 
Many predictions related to the gravitational-wave background (GWB) produced by inspiralling MBHs have been made 
(e.g., \citet{Sesana2008, Ravi2012, Kelley2017_ills_env, Izquierdo-Villalba2022, Sah2024, Saeedzadeh2024}). 
However, they all predict a GWB amplitude smaller than that measured observationally. 
\citet{Sato-Polito2024_diff} and \citet{Sato-Polito2024_dist} demonstrated that even with the optimal scenario for the MBH merging history and accretion, the difference between predictions and observations is still significant, and the current GW measurement requires roughly $\sim$ 10 times more black holes than suggested by local observation.

\begin{figure*}[!ht]
\centering
	\includegraphics[width=0.8\textwidth]{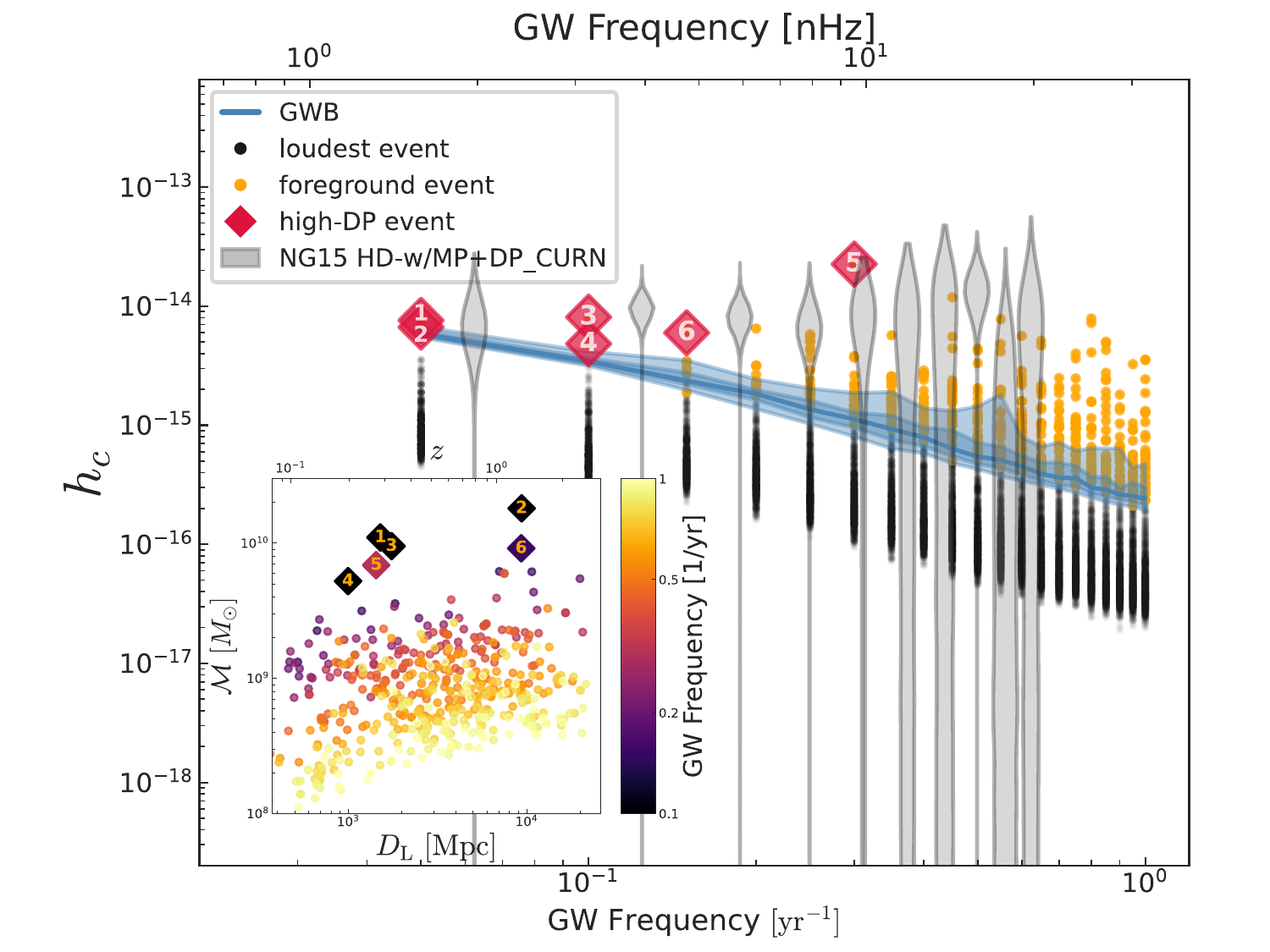}
    \caption{
    $100$ \reviseagain{independent} realizations of the low-frequency GW sky based on the mergers in \astrid. The blue curve represents the median GWB spectra over all realizations, with the inner/outer shaded region corresponding to $1\sigma$/$2\sigma$ intervals.  
    Black dots plot the loudest event in each realization at each frequency bin. The foreground events, which have higher $h_{\mathrm{c}}$ than the GWB are marked by orange dots. The grey shaded area is the NANOGrav 15Yr GWB. We highlight the events with high detection probability (with DP $> 0.1$) using red diamonds.
    The insert shows the distribution of the chirp mass $\mathcal{M}$ and the luminosity distance $D_{\mathrm{L}}$ for all foreground sources, color-coded by the observed frequency. The high-DP events, marked by the same index as the GWB plot, are plotted by diamonds. 
    }\label{fig:GWB}
\end{figure*}

Continuous waves (CWs) from signal loud MBH binaries 
are the next anticipated GW signal \citep{Rosado2015, Kelley2018_SS, Mingarelli2017}. 
The CW signals are also known as GW foregrounds; hereafter we use them interchangeably. 
Before the CWs can be individually resolved, the anisotropy they introduced in the low-frequency sky might be detected \citep{Becsy2022}. 
Despite the non-detection in the search for either individual MBH binaries or anisotropy in the NANOGrav 15 yr Data Set \citep{Agazie2023_NANOgrav_SSserch, Agazie2023_anisotropy_search}, some simulation-based predictions for CW sources have been made by previous works.
Using the galaxy catalog from the 2 Micron All Sky Survey \citep[2MASS;][]{Skrutskie2006_2mass}, together with galaxy merger rates from the 
cosmological simulation Illustris \citep{Genel2014_illustris, Rodriguez-Gomez2015_illu_mergingrate},
\citet{Mingarelli2017} estimated that there are on average $91\pm 7$ detectable CW sources within 225 Mpc over the full sky, and a 20\% departure from an isotropic GWB induced by local unresolved sources. 
Utilizing the MBH binary population from the Illustris, \citet{Kelley2018_SS} calculated the plausible detection prospects for GW single sources and predicted that they are as detectable as the GW background. 
\citet{Gardiner2024_PTACW} investigated the detectability of CWs 
over a wide range of parameter space, including the binary evolution prescription. They predict that the most detectable CW sources are in the lowest frequency bin for a 16.03-yr PTA, having masses from $10^{9}\sim 10^{10}$ \Msun\ and are $\sim$ 1 Gyr ($z\sim0.2$) away. 

In this work, we present a prediction of the CW sources detectable by PTAs based on the MBH merger catalog from the cosmological simulation \astrid.
This paper is organized as follows. In Section~\ref{sec:methods} we introduce the simulation \astrid\ and the methods used to simulate single GW events. 
In Section~\ref{sec:results}, we present our CW prediction, and provide a detailed analysis of the properties of the high-detectability sources. Finally, we conclude in Section~\ref{sec:conclusion}.

\section{Methods}
\label{sec:methods}
In this work, we use the MBH merging population from the \astrid\ cosmological hydrodynamic simulation to model the low-frequency GW sky.
\astrid\ is the largest cosmological hydrodynamical simulation, in terms of particle load, so far run to $z=0$. 
It contains $2 \times 5500^{3}$ particles in a box $250h^{-1}$Mpc per side, where $h=0.6774$. 
Here we briefly introduce the basic parameters and BH modeling for \astrid, and refer readers to \citet{Ni2022_astrid,Bird2022_astrid,Ni2024} for more details. 
The mass resolution of \astrid\ is $m_{\mathrm{DM}} = 6.74 \times 10^{6}\,h^{-1}\mathrm{M}_{\odot}$ and $m_{\mathrm{gas}} = 1.27 \times 10^{6}h^{-1}\mathrm{M}_{\odot}$. The gravitational softening length is $\epsilon_{\mathrm{g}}=1.5 h^{-1}$ kpc for all  particles.
MBHs are seeded with their mass stochastically drawn from $3\times 10^{4}h^{-1}\ \mathrm{M}_{\odot}$ to $3\times 10^{5}h^{-1}\ \mathrm{M}_{\odot}$.
The gas accretion rate of black holes is given by the Bondi-Hoyle rate, and super-Eddington accretion is allowed with an upper limit of twice the Eddington accretion rate.
With a radiative efficiency $\eta=0.1$ \citep{Shakura1973_BH}, the black hole radiates with a bolometric luminosity $L_{\mathrm{bol}}$ proportional to the accretion rate: $L_{\mathrm{bol}}=\eta\,\dot{M}_{\mathrm{BH}} c^{2}$. Both AGN thermal and kinetic feedback models are included.
A subgrid dynamical friction model is used based on the prescription in \citet{Tremmel2015_DFmodel, Chen2022_DFmodel} to capture the dynamics of MBHs as they merge. 
\revise{This model has been validated against both semi-analytical predictions and high-resolution simulations \citep{Genina2024}.
For example, \citet{Zhou2025} resimulated several merging systems in \astrid. With a higher resolution and the DF resolved self-consistently, the binary evolution agrees well with that predicted by the DF subgrid model on the scales covered by \astrid.  
} The dynamical friction significantly affects the MBH binary populations and predictions for the GWB \citep{PTA_astrid_GWB}. 
Two black holes merge when their separation is within $2\epsilon=3\mathrm{ckpc}/h$ and they are gravitationally bound to the local potential. 

\reviseagain{After the DF subgrid model evolves the MBH binaries to $\sim 1$~kpc scales, other mechanisms, like loss-cone scattering and gas drag from the circumbinary disk, are required to further decay the MBH orbits to sub-parsec scales, where they can eventually emit the GW signals detected by PTA.
To account for the hardening on the scales of $<1$~kpc, which is unresolved in \astrid,} 
we follow \citet{PTA_astrid_GWB} and use the 
phenomenological hardening model in \texttt{Holodeck}\footnote{\url{https://github.com/nanograv/holodeck}}. 
We use the same parameter settings as  \citet{PTA_astrid_GWB}, which are based on the best-fit value of the \textit{Phenom+Astro} analysis in \cite{Agazie2023_GWB_smbh_constrain}. Specifically, the binary hardening timescale is fixed to be $\tau=500$~Myr.
\reviseagain{On the one hand, adopting a larger $\tau$ will move the entire MBH binaries population to lower redshifts, boosting the detected strain for individual sources. 
On the other hand, the large hardening timescale makes massive sources less likely to reach small enough separations to emit at the PTA band by $z=0$, resulting in a decrease in the number of detectable sources. 
According to \citet{Gardiner2024_PTACW}, the detection probability for CW sources increases by a factor of five when $\tau$ changes from 0 Gyr to 1 Gyr, and then drops slowly for larger $\tau$. 
}

\reviseagain{
In this work, we assume circular orbits for all the binaries during the evolution. 
The eccentricity of MBH bound binaries in the GW regime is notoriously difficult to predict, which can be sensitive to the initial orbital eccentricity and the galactic environment \citep{Sesana2010, Roedig2012,
Fastidio2024, Franchini2024}.
An accurate estimation for the PTA signals from MBH binaries in eccentric orbits is out of the scope of this letter, and we will explore this in future work. 
Some previous studies incorporated the eccentricity in their predictions for the PTA CW signals, while achieved different conclusions: \citet{Kelley2018_SS} showed that incorporating eccentricity would decrease the foreground occurrence rate, while \citet{truant24} found that larger initial eccentricity makes more CW sources resolvable. However, both of these works found that the binary intrinsic properties of the CW sources do not depend on the eccentricity except for the extreme case $e = 0.99$. Hence, although the circular assumption simplifies our predictions in this work, our results related to the properties of detectable CW sources are robust.} 
Given the evolved population of binaries, the production of GW signals is estimated based on the steps laid out by \citet{Sesana2008}. To create discrete realizations of GWB sources from simulation events, we use the method described in \citet{Kelley2017_ills_env} to weight each MBH binary from \astrid\ using a Poisson distribution. We refer the readers to Appendix A and Appendix B in \citet{PTA_astrid_GWB} for more details related to this part. 

With the strain of each source, we calculate the detection probability (DP) and the expected number $\left<N_{\mathrm{SS}}\right>$ of detected single sources based on the prescription of \citet{Rosado2015}.  We present the details about the adopted pulsar noise model in Appendix~\ref{app:DP}.


\section{Results}
\label{sec:results}
\subsection{Characteristic Strain and DP}
\label{sec:results:strain_DP}

Drawing on the MBH mergers in \astrid\ down to redshift $z=0$,
we generate 100 \reviseagain{independent} realizations of the low-frequency GW sky covering the frequency range $0.05/\mathrm{yr} < f< 1/\mathrm{yr}$ (i.e., $1.6\sim 32$ nHz) with a timing cadence $\Delta t = 0.05$ yr. 
The results are presented in Fig.~\ref{fig:GWB}.
The blue curve represents the median GWB across all  realizations, and the inner/outer shaded region shows the 1$\sigma$/$2\sigma$ intervals. 
Black dots show the loudest event in each realization at each frequency bin, and orange dots represent the foreground events.
In this work, we define the foreground events as those having higher strain $h_{\mathrm{c}}$ than the GWB in the same realization.
Due to the low GWB at high frequencies, there are more foreground events at $f\sim1$/yr compared to low frequencies. 
At $f<0.1$/yr, all the realizations include loud mergers with $h_{\mathrm{c}}>10^{-16}$, while only 5 of them make foreground sources.
In the insert of Fig.~\ref{fig:GWB}, we show the chirp mass $\mathcal{M}$ versus the luminosity distance $D_{\mathrm{L}}$ of all the foreground events, color-coded by their observed frequency. The correlation between $\mathcal{M}$ and $D_{\mathrm{L}}$ is not obvious, except that the lower bound of $\mathcal{M}$ extends to smaller values with smaller $D_{\mathrm{L}}$, i.e., at lower redshift, we are expected to detect less massive CW sources.  
The presented color gradient in the points is clearly seen, however, showing that how low chirp mass events appear in higher frequency bins. 

\begin{figure*}[!ht]
\centering
    \includegraphics[width=\textwidth]{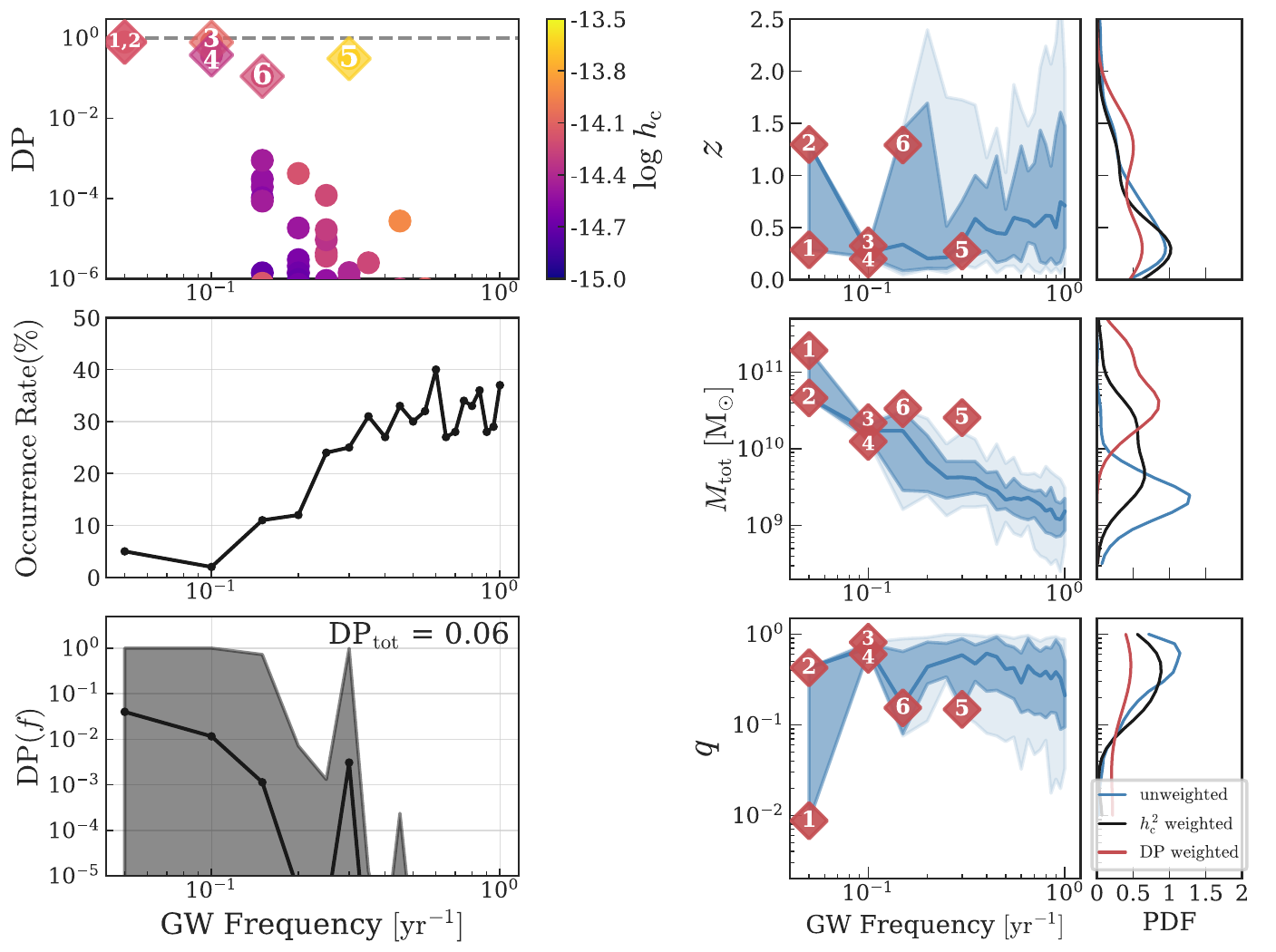}
    \caption{ The properties of foreground events over 100 realizations. \textit{Left Column}:
    the top panel shows the DP values for each CW source color-coded by $h_c$.
    We highlight the sources with DP$> 0.1$ (high-DP events) by diamonds. \reviseagain{The diamond markers for system 1 and system 2 overlap.}
    The middle/bottom panels show the occurrence rate / DP  of the detected CW sources 
    as a function of frequency. 
    In the bottom panel, the grey shaded areas are the $1\sigma$ region over 50000 settings (100 GW sky realizations $\times$ 500 sky-position settings; see Appendix~\ref{app:DP}). 
    \textit{Right Column}:
    From top to bottom, we show the redshift $z$, 
    the total mass of the merging BHs ($M_{\mathrm{tot}} = M_{\mathrm{BH,1}} + M_{\mathrm{BH,2}}$),
    and the mass ratio $q=M_{\mathrm{BH,2}}/M_{\mathrm{BH,1}}$ for all the CW sources at each frequency. 
    In each panel, the median (blue curves), 1$\sigma$ intervals (dark shaded area), and $2\sigma$ intervals (light shaded area) are plotted. In the right frame of each panel, we show the unweighted (blue), $h_{\mathrm{c}}^{2}$-weighted (black), and DP-weighted (red) PDF. 
    }\label{fig:binary_DP}
\end{figure*}

\begin{table}[!ht]
\centering
\caption{The 6 CW sources with DP $> 0.1$}
\begin{tabular}{cccccccc}
\specialrule{0.08em}{0.05pt}{2pt}     
\specialrule{0.08em}{0.05pt}{2pt}     
      & $z_{\mathrm{merge}}$
      & $M_{\mathrm{BH,\, 1}}$  
      & $q$
      & SFR$^{\rm \ast}$
      & $M_{\mathrm{\star, gal}}$ $^{\rm \dagger}$
      & $M_{\mathrm{FOF}}$ 
      \\ 
      & 
      & $10^{10}\mathrm{M}_{\odot}$  
        & 
        & $\mathrm{M}_{\odot}/\mathrm{yr}$
      & $10^{12}\mathrm{M}_{\odot}$  
      & $10^{14}\mathrm{M}_{\odot}$  
      \\ \specialrule{0.08em}{4pt}{4pt}
sys 1 & 0.29 & 20.0 & 0.01 & \reviseagain{5} & \revise{2.5} & 6.3 \\
sys 2 & 1.30 & 3.2 & 0.43 & \reviseagain{20} & \revise{2.8} & 5.0 \\
sys 3 & 0.32 & 1.3 & 0.81 & \reviseagain{75} & \revise{5.0} & 15.8 \\
sys 4 & 0.20 &  0.8 & 0.60 & \reviseagain{8} & \revise{1.4} & 4.0 \\
sys 5 & 0.27 & 2.5 & 0.15 & \reviseagain{72} & \revise{5.6} & 15.8 \\
sys 6 & 1.29 & 3.2 & 0.15 & \reviseagain{8} & \revise{1.2} & 2.5 \\
\bottomrule
\end{tabular}\label{tab:high-DPsources}
    \begin{justify}
    \reviseagain{$^{\rm \ast}$ The SFR is calculated as the averaged star formation rate over the past 50 Myrs within a fixed aperture of 20~kpc.\\
    $^{\rm \dagger}$ $M_{\star, \rm gal}$ is the stellar mass within twice the stellar half-mass radius.  
    }
    \end{justify}
\end{table}

\begin{figure*}[!ht]
	\includegraphics[width=\textwidth]{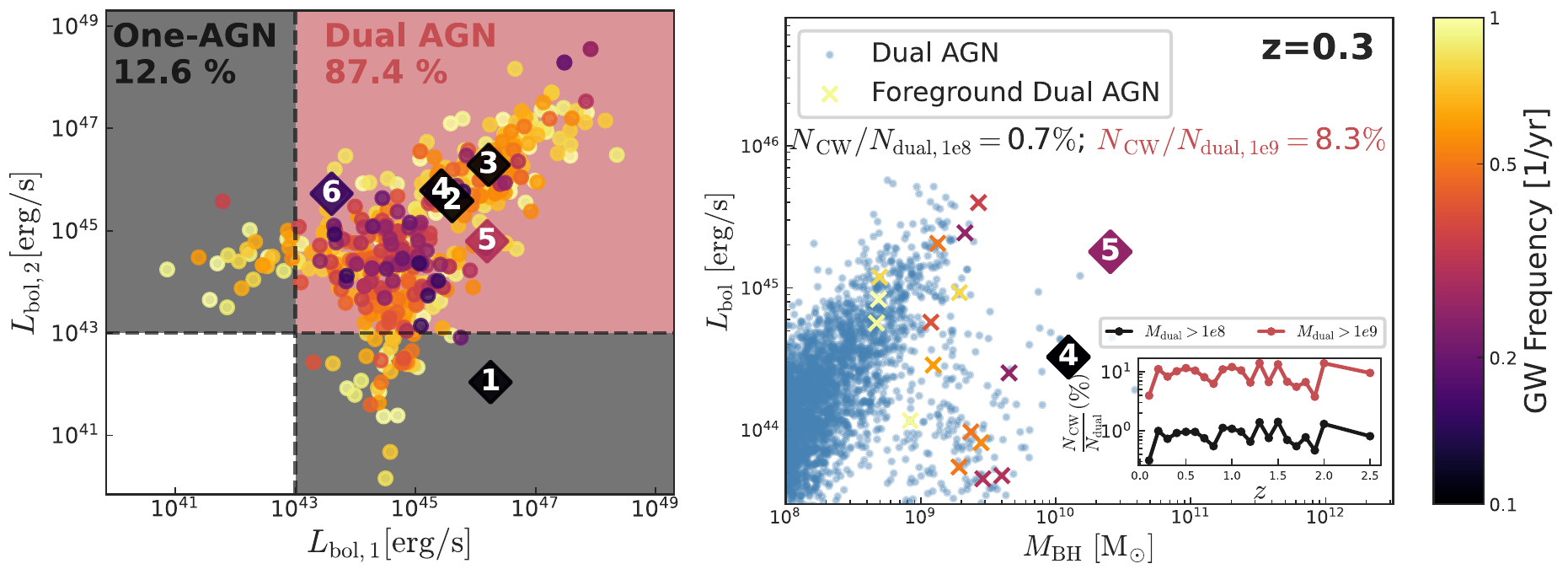}
    \caption{
\textit{Left}: The distribution of the bolometric luminosity of the primary MBH ($L_\mathrm{bol,1}$) and the secondary MBH ($L_\mathrm{bol,2}$) for the foreground event population. 
    The dots are color coded by the observed frequency. 
    The horizontal and vertical black dashed lines mark the AGN luminosity limit: we define an MBH with $L_{\mathrm{bol}} > 10^{43}\ \mathrm{erg}/s$ as an AGN. The red shaded area on the upper right corner represents the dual AGN regions, where both MBH in the merging pair are AGNs. Dual AGN comprise 87.4\% of the foreground events. 
    The two black shaded areas represent one-AGN regions, for which only one of the MBHs is an AGN, and contain 12.6\% of the foreground events.     
\textit{Right}: The distribution of the bolometric luminosity $L_{\mathrm{bol}}$ and the merging MBH mass $M_{\mathrm{BH}}$ for the dual AGNs at $z=0.3$. 
    The blue dots are the whole dual AGN population in \astrid\ at the specific redshift. 
    The crosses are the events among the dual AGNs that evolve to be foreground events within 1 Gyr, which are color coded by the observed GW frequency.
    The two panels share the color bar on the right. 
    For the dual AGN with a mass over $10^{8}\,\mathrm{M}_{\odot}$, about 0.7\% of them will evolve to CW sources. For dual AGN with  $M_{\mathrm{BH}}>10^{9}\,\mathrm{M}_{\odot}$, $N_{\mathrm{CW}}/N_{\mathrm{dual}}$ increases to 8.3\%. 
    We show the evolution of $N_{\mathrm{CW}}/N_{\mathrm{dual}}$ in the inserted frames: black (red) curve represents the CW fraction for dual AGN with the mass above $M_{\mathrm{BH}}>10^{8}\,(10^{9})\,\mathrm{M}_{\odot}$
    }
    \label{fig:DUAL_frac}
\end{figure*}

We calculate the DP for all the foreground sources using the prescription described in Appendix~\ref{app:DP}. 
The results are shown in the left column (top panel) of Fig.~\ref{fig:binary_DP}.
If one merger is observed as a foreground event at the same frequency in different realizations, we only show it once with the highest DP. 
\reviseagain{Among these 100 independent realizations, there are only 6 sources that have DP higher than 0.1, and they are from six different realizations.}
We mark them as ``high-DP sources'', and show their positions among the foreground distribution in Fig.~\ref{fig:GWB}, \ref{fig:binary_DP}, \ref{fig:DUAL_frac}, and \ref{fig:hostgal}. 
All the high-DP sources are among the lowest frequencies with $f\lesssim 0.3$/yr ($\sim 10$ nHz). The most likely frequency of detection is predicted by estimating the DP-weighted frequency across all mergers: $ \left<f_{\mathrm{fore}}\right> = {\sum_{i}\mathrm{DP}_{i}\,f_{i}}/{\sum_{i}\mathrm{DP}_{i}}=0.07$/yr, where $\mathrm{DP}_{i}$ and $f_{i}$ are the DP and frequency for a specific source. 
This is consistent with the estimation made in \citet{Gardiner2024_PTACW}.
The DP drops rapidly with increasing frequency, and there are no mergers with $\mathrm{DP}>10^{-6}$ with $f>0.5$/yr.

The second panel in the left column of Fig.~\ref{fig:binary_DP} shows the foreground occurrence rate, i.e., the fraction of the 100 realizations that host at least one CW sources in the specific frequency bin. Due to the relatively low GWB at the high-$f$ end, more mergers can be resolved as CW, leading to higher occurrence rates. Our prediction result is generally consistent with \cite{Kelley2018_SS}.
The bottom panel shows the probability of detecting at least one CW source at given $f$: DP$(f)$.
The black curve represents the median among all the realizations, and the shaded area marks the whole range. 
Although the occurrence rate at the low-$f$ end is very small ($\sim 5\%$ at $f<1$/yr) compared to the high-$f$ end ($\gtrsim 20\%$ at $f>1$/yr), DP at low-$f$ outweighs that at high-$f$. This is because all the CW sources at $f\lesssim1$/yr 
have very high detectability, as shown in the top panel of the column.
Since the lowest frequency is determined by the observation duration $T$, our results imply that increasing $T$ would significantly boost the chance to observe CW sources. 

The probability for an individual realization to detect at least one CW source across the whole frequency range is 
$\left<\mathrm{DP}_{\mathrm{tot}}\right>=0.06$. 
\revise{This is not only consistent with the predictions made based on the Illustris cosmological simulation \citep{Becsy2022}, but also aligned with the non-detection of the CW sources in the NANOGrav 15yr Data Set \citep{Agazie2023_NANOgrav_SSserch}. 
This low detection probability implies the necessity to include more pulsars and to improve the pulsar noise models \citep{Kelley2018_SS, Goncharov2021}. 
}


\revise{
}

\begin{figure*}[!ht]
\centering
	\includegraphics[width=0.9\textwidth]{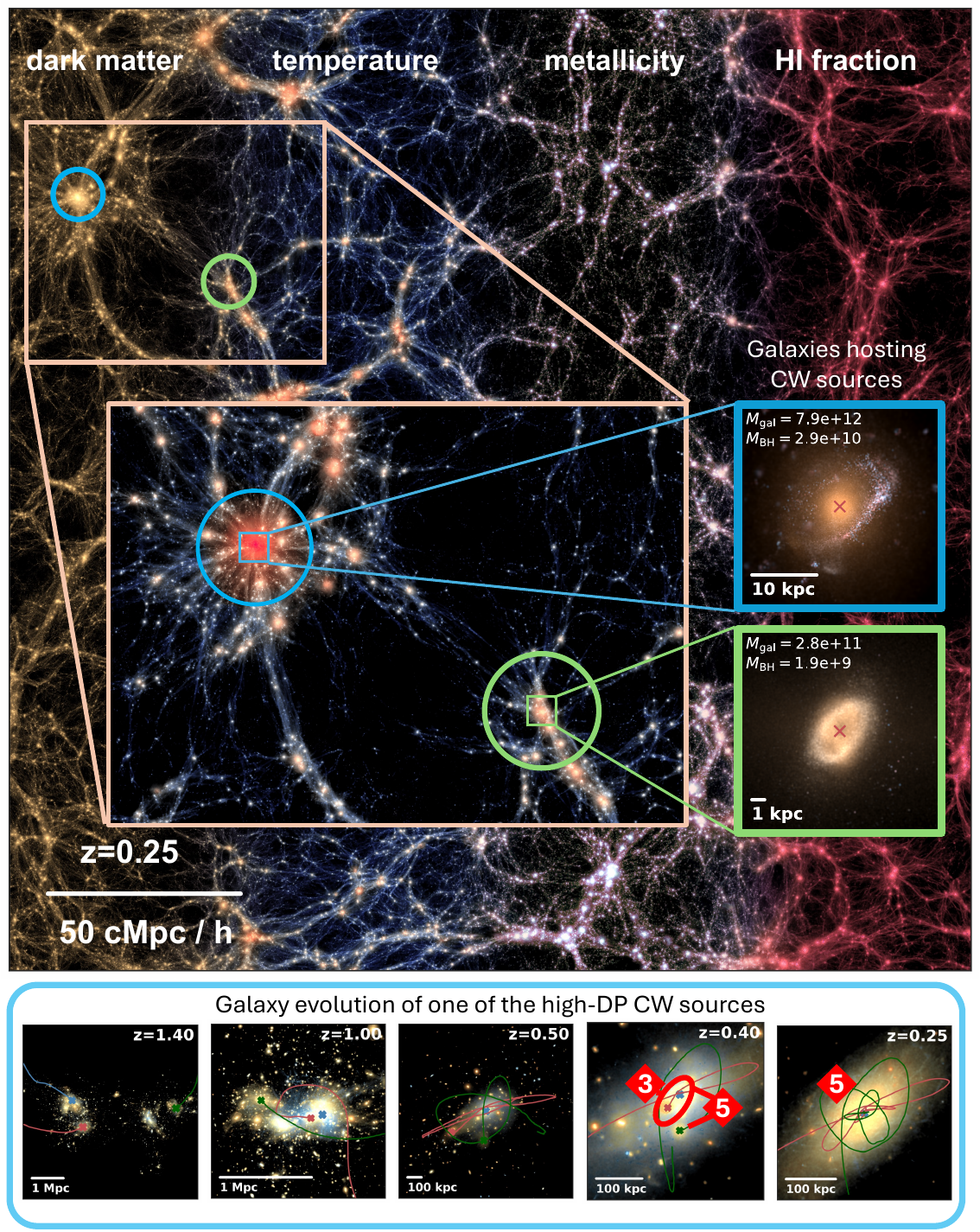}
    \caption{  A visualization of the spatial region around the high-DP event system 3 and system 5 at $z=0.25$.
    These two systems involve a triple merger in the central galaxy of a massive galaxy cluster, \reviseagain{which is highlighted by the blue circles.}
    We zoom into a region around it \reviseagain{(the orange rectangle)}, 
    which includes another foreground event \reviseagain{(marked by the green circle)}, and show the gas density field color-coded by temperature (red color represents hotter gas). 
    \reviseagain{The two small insets on the right show the $grz$-band flux of the host galaxies of the two CW sources, which are located at the center of the blue/green halos. The position of the merging remnant MBHs is marked by red crosses.}
    The panels in the bottom row illustrate the orbital evolution of the triple merger in system 3 and 5.
    In each frame, the blue/red/green crosses mark the position of the primary/secondary/third MBH. Their trajectories are plotted with corresponding colors. 
    The black holes involved in the events that are part of the triple merger  (system 3 and then system 5) are indicated in the bottom right two panels. 
    In the frames with $z\leq1$, the trajectories are plotted relative to the primary MBH. 
    The underlying field is the star density color-coded by the star age, with blue to yellow representing young to old stars. 
    }
    \label{fig:largescale}
\end{figure*}

\subsection{MBH pair properties}

To further explore the frequency dependence of the CW sources, in the right column of Fig.~\ref{fig:binary_DP} we display the redshifts $z$ (upper panel), 
masses of merging MBH $M_{\mathrm{tot}} = M_{\mathrm{BH,1}} + M_{\mathrm{BH,2}}$ (middle panel), and MBH mass ratios $q = M_{\mathrm{BH,2}}/M_{\mathrm{BH,1}}$ (bottom panel) for all the foreground sources. $M_{\mathrm{BH,1}}$ and $M_{\mathrm{BH,2}}$  are the mass of the primary and secondary MBH, respectively. 
In each panel, blue curves show the median value, and the dark (light) shaded areas are the $1\sigma$ ($2\sigma$) regions. 
As we have seen from the insert panel in Fig.~\ref{fig:GWB},  merger mass is strongly correlated with the observed frequency.
At low frequency, the foreground population is dominated by the massive mergers: all the sources observed at $f<0.1$/yr are above $10^{10}\ \mathrm{M}_{\odot}$.
However, we see that neither redshift nor mass ratio show any obvious dependence on the observed $f$. 
 
We plot the one-dimensional probability distribution function (PDF) for $z$, $M_{\mathrm{tot}}$,
and $q$ in the attached panel of the right column.
The blue, black, and red curves represent the unweighted, $h_{\mathrm{c}}^{2}$-weighted, and $\mathrm{DP}$-weighted PDF. 
Examining the PDF of $z$, we see that a large fraction of the foreground sources occurs below $z<1$, and the peak for the unweighted PDF is at $z\sim0.3$ ($D_{\mathrm{L}} \sim 1600\ \mathrm{Mpc}$). 
After weighting by the GW energy ($h_{\mathrm{c}}^{2}$), the peak moves even lower to $z\sim0.2$. This is expected since low-$z$ sources generally produce higher $h_{\mathrm{c}}$. 
The PDF weighted by DP, however, is double-peaked, with peaks at low redshift $z\sim0.2$ and around $z=1.2$. The latter peak is caused by the two high-DP events: system 2 and system 6.
The drop of the CW sources number at low redshift ($z<0.2$) is due to the decreasing  MBH merger rate: 
as showed in \citet{PTA_astrid_GWB}, the merger rate of massive MBHs (with $M_{\mathrm{tot}} \geq 10^{9}\ \mathrm{M}_{\odot}$) in \astrid\ starts to drop below $z<0.5$, so there are more supermassive mergers to be detected at high redshift rather than in the local Universe ($z\sim0$). 
An individual example of such a massive event could change the shape of the DP-weighted $z$ distribution, which highlights that PTA has a large probability of observing mergers with $z>1$.    
From the bottom panel in the right column of Fig.~\ref{fig:binary_DP}, we can see that most CW sources are major mergers with $q\gtrsim 0.1$, and the median $q$ is around 0.5. 
One exception to this is system 1, which has a very low mass ratio $q=0.01$ and involves the most massive MBH in \astrid: $M_{\mathrm{BH}}=2\times 10^{11}\ \mathrm{M}_{\odot}$.



\subsection{High-DP CW sources}
\label{sec:samplestudy}
In this section, we present more details about the 6 high-detectability CW sources with DP $> 0.1$. In Table~\ref{tab:high-DPsources}, we list the properties of the merging MBHs, and the masses of their host galaxy and Friend-Of-Friends (FOF) halo.  
These six sources are all from the high mass end among the merger population in \astrid, with the total MBH masses $M_{\mathrm{BH}} >10^{10}\ \mathrm{M}_{\odot}$. However, they cover a wide range in redshift ($0.3<z<1.3$) and mass ratio ($0.01\lesssim q\leq1$).
All these sources are located in the massive galaxies with $\log M_{\mathrm{\star, gal}} > 10^{12}\ \mathrm{M}_{\odot}$. 
These host galaxies are all the central galaxies of the massive galaxy clusters with the total masses larger than $10^{14}\,\mathrm{M}_{\odot}$. 
This highlights that central cluster galaxies would be a hotspot for detectable CW sources. 
\reviseagain{We want to remind the reader that our prediction is made based on all the MBH mergers in \astrid\, without imposing any mass cut or other selection criteria on the host galaxy. Hence, the massive galaxy hosts are a natural emergence in our estimation.}

Remarkably, among these 6 high-DP sources, 
system 3 and system 5 are involved in a triple merger event, wherein two consecutive mergers happen within a 500 Myr interval in the same cluster core. 
System 3 is the first merger, which involves MBHs with masses $1.2\times 10^{10}\ \mathrm{M}_{\odot}$ and $9.9\times 10^{9}\ \mathrm{M}_{\odot}$  at $z=0.32$. It is observed at $f=0.1$/yr ($\sim3$ nHz) as a CW source with DP=0.89.
System 5 is the second merger that occurs at $z=0.27$, and involves a $2\times 10^{10}\ \mathrm{M}_{\odot}$ MBH and a $3\times 10^{9}\ \mathrm{M}_{\odot}$ MBH.  It is observed at $f=0.3$/yr ($\sim10$ nHz) with a DP of $0.3$. 
System 5 happens 480 Myrs after system 3. This timescale is slightly shorter than our adopted binary hardening timescale $\tau=500$~Myr, implying the possibility that this triple-merger could form an actual three-body system.
\reviseagain{Fig.~\ref{fig:largescale} shows a visualization of the spatial region around this triple merger, marked by the blue circle, at $z=0.25$. We zoom into the region around it (the orange rectangle), 
which includes another foreground event with however has very low DP, which is highlighted by the green circle.}
We show the gas density field color-coded by temperature, where the red color represents the hot region. Additional small inserts show the $grz$-band flux of the host galaxies of the two CW sources, with the merging remnant MBH marked by red crosses.
The host galaxy of the remnant of the triple merger has a stellar mass of $M_{\mathrm{\star}}=7.9\times10^{12}\,\mathrm{M}_{\odot}$.
The orbital evolution of the triple merger is shown in the lower panels in Fig.~\ref{fig:largescale}, starting at $z=1.2$ and continuing until the end of the triple merger (z=0.25).
Following the sequence over this 6 Gyr period, we can see the formation of the most massive galaxy in this cluster. At $z=1.2$ several distinct galaxies are visible, most of which merge over the next  4 Gyr. High-DP CW event 3 occurs in the center of the brightest cluster galaxy (BCG), and then a smaller galaxy containing the last of three MBHs falls into the BCG
in an act of galactic cannibalism (e.g., \citealt{hsu22}). This is followed by the final 
merger, high-DP source 5.
 


\subsection{Dual AGN}
In this section, we study the correlation between CW events and dual AGN, which has an important implication for future searches for the EM counterpart of GW sources. 
We use the same definition of dual AGN as \citet{Chen2023_dualAGN}: if the MBH bolometric luminosity $L_{\mathrm{bol}}>10^{43}\ \mathrm{erg}/s$, the MBH is classified as an AGN. 
A pair of AGNs is classified as dual AGN if the separation between them is smaller than $30$ kpc. To focus on the same mass range as the PTA CW sources, we only study dual AGNs with total $M_{\mathrm{BH}}$ larger than $10^{8}$ \Msun.

In the left panel of Fig.~\ref{fig:DUAL_frac}, we show the distribution of the bolometric luminosity of the primary MBH ($L_\mathrm{bol,1}$) versus that of the secondary MBH ($L_\mathrm{bol,2}$) for the CW sources population. 
In the upper right, both MBH are bright enough that the event is categorized as coming from a dual AGN. This encompasses the vast majority (87.4\%) of the foreground events.
All the other CW sources (12.6\%) are ``One-AGN pair'', in which one of the merging MBH is AGN, while the other is not. 
Hence, all the CW sources predicted by \astrid\ in the frequency range $0.05/\mathrm{yr}<f<1/\mathrm{yr}$ are involved in AGN activity.

In the right panel of Fig.~\ref{fig:DUAL_frac}, we further investigate the dual AGN population.  Looking at the luminosity $L_{\mathrm{bol}}$ versus MBH mass of dual AGN in \astrid\ at $z=0.3$, we can see that the CW dual AGN sources (marked by crosses) cover the whole luminosity range, but they preferentially pick out BHs at the high mass end.
Among the dual AGN with $M_{\mathrm{tot}} \geq 10^{8}\ \mathrm{M}_{\odot}$ at $z=0.3$,
only 0.7\% will evolve to become foreground sources. 
While for the dual AGNs with $M_{\mathrm{tot}} \geq 10^{8}\ \mathrm{M}_{\odot}$, the CW sources fraction $N_{\mathrm{CW}}/N_{\mathrm{dual}}$ increases to 8.3\%.
Three of the six high-detectability CW sources, system 1, 4, and 5, are detected at $z\leq0.3$. As shown in the left panel, system 1 involves a `one-AGN pair' MBH. Hence, we only mark system 4 and system 5 with diamonds in the right panel. 
In the insert frame in the right panel, 
we show the evolution for $N_{\mathrm{CW}}/N_{\mathrm{dual}}$.
We can see that the fraction of dual AGNs that evolve into CW sources does not change with redshift and stays at $\sim 1 \%$ ($\sim 10\%$) for the mass range $M_{\mathrm{BH}}>10^{8}\ \mathrm{M}_{\odot}$ ($M_{\mathrm{BH}}>10^{9}\ \mathrm{M}_{\odot}$).

\subsection{Host Galaxies of CW sources} \label{sec:hostgal}

\begin{figure*}
	\includegraphics[width=\textwidth]{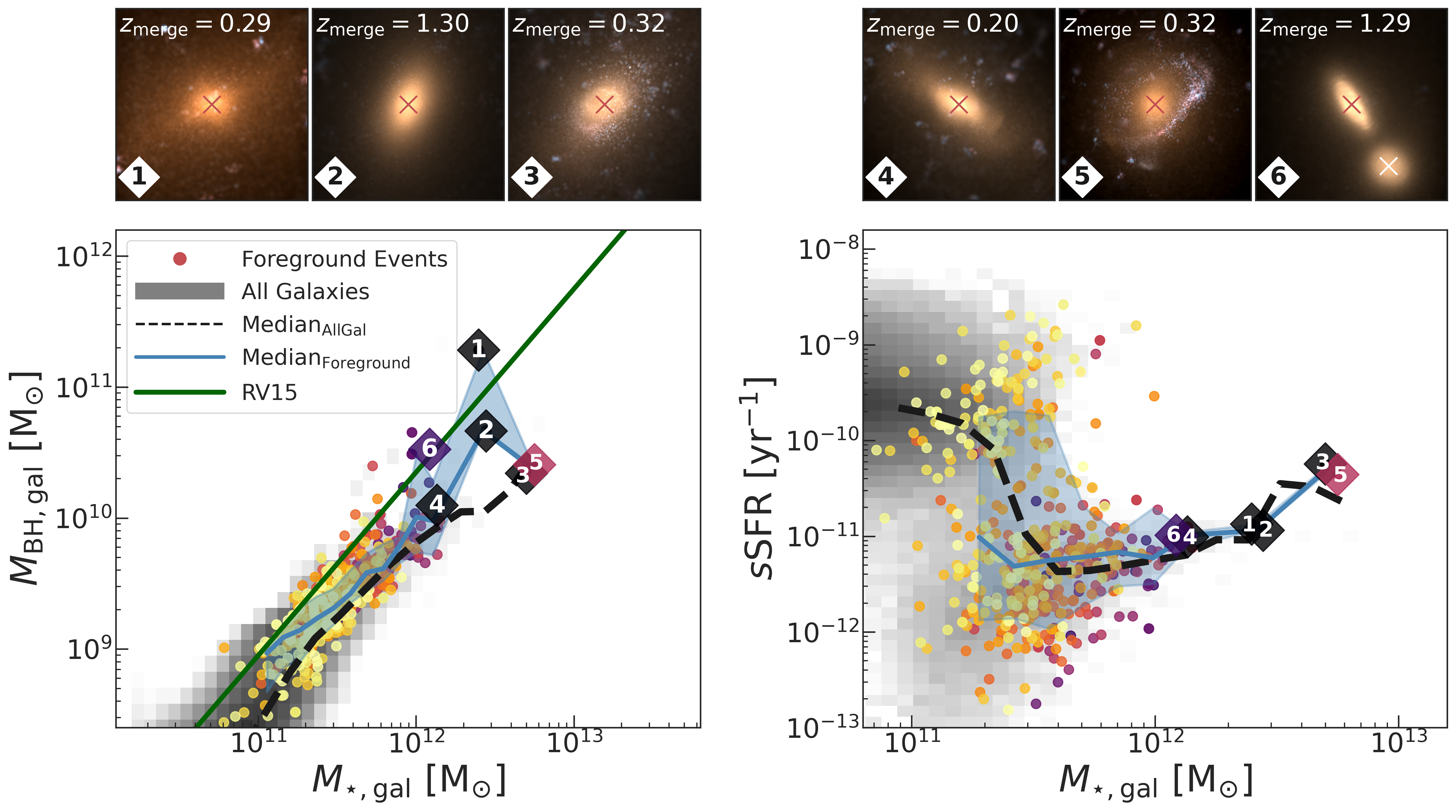}
    \caption{
    The host galaxy properties of the foreground events (colored dots) compared to all the galaxies in \astrid\ (grey pixels). 
    The underlying darker regions represent a higher number density. 
    The dots in two panels share the same colorbar as Fig.~\ref{fig:DUAL_frac}.
    \textit{Upper:} the projected figures of the host galaxies of the six high-DP CW sources, with the RGB channels representing the flux in the rest-frame $grz$ color bands. 
    \textit{Bottom Left:} the distribution of the stellar mass and the \revise{central} black hole mass in the galaxies. 
    To guide the eye,
    we add the scaling relation drawn from the observation population in \citet{Reines2015}.
    \textit{Bottom Right:} the distribution of the specific star formation rate $s$SFR versus the stellar mass in the galaxies. 
    In both panels, the black dash/blue solid curves show the median values for all galaxies/foreground sources. 
    \reviseagain{The $M_{\star,\rm gal}$ is the stellar mass within twice the stellar half-mass radius. The specific star formation rate $s$SFR is measured within a fixed aperture of 20 kpc and averaged over the past 50 Myrs.}
    }
    \label{fig:hostgal}
\end{figure*}

In this section, we investigate the host galaxy of GW foreground events. 
We identify the galaxy hosting the remnant of each CW source in the closest snapshot following the merger. 
The upper frames in  Fig.~\ref{fig:hostgal} show 
the projected figures of the host galaxies of the six high-DP CW sources, with the RGB channels representing the flux in the rest-frame $grz$ color bands. 
The central red crosses mark the position of the remnant MBH. 
For system 6, one of the high-redshift sources at $z=1.29$, the host galaxy is involved in a merger with another galaxy, and we mark the other central MBH with a white cross. 
In the bottom two panels we show the $M_{\mathrm{\star,gal}}$ - $M_{\mathrm{BH}}$  (left) and specific star formation rate ($s$SFR) versus the galaxy stellar mass (right) of the 
foreground events (colored dots) compared to all the galaxies in \astrid\ in $0<z<7$ (underlying grey distribution).
For $M_{\mathrm{\star,gal}}$ - $M_{\mathrm{BH}}$ panel, \revise{we use the mass of the galaxy's central BH as $M_{\rm BH}$.}
To guide the eye, we add the scaling relations drawn from the observation population in \citet{Reines2015}.
\revise{In both panels, the $M_{\star,\rm gal}$ is the stellar mass within twice the stellar half-mass radius.} \reviseagain{The specific star formation rate $s$SFR is measured within a fixed aperture of 20~kpc and averaged over the past 50 Myrs.}
We can see that the foreground sources follow a trend consistent with that of the whole galaxy population. The median relation for CW sources closely tracks the relation for all galaxies across the whole $M_{\mathrm{\star,gal}}$ range. The individual high-DP sources do not differ significantly from the mean relation, except for source 1, which appears to be an overmassive MBH.
The events arising from a triple merger, system 3 and system 5, are associated with the most massive end of the relation.
\revise{
The six high-DP sources all have a $s$SFR  $\gtrsim 10^{-11}$/yr. 
In Table~\ref{tab:high-DPsources}, we list the total SFR for each event. These galaxies all have a SFR larger than 5 $\mathrm{M}_{\odot}/\mathrm{yr}$.
Among them, the events related to the triple merger have a higher $s$SFR compared to other high-DP sources, whose total SFR is over 70 $\rm M_{\odot}/{\rm yr}$. This is likely due to the recent intense galaxy merger activity.
}

\section{Discussion and Conclusions}
\label{sec:conclusion}
This work, using the ASTRID cosmological hydrodynamic simulation, provides significant new insights into the nature of detectable CW sources for 16.03yr-PTAs. Our analysis, focusing on supermassive black hole (SMBH) mergers, reveals several crucial findings:

$\bullet${\it Central Cluster Galaxies as a Hotspot}: We find that the most detectable CW signals are produced by MBH mergers with masses exceeding $10^{10}\ \mathrm{M}_{\odot}$, predominantly located within the massive central galaxies of galaxy clusters. This highlights a previously under-appreciated connection between these high-mass CW sources and the formation process of central cluster galaxies.

$\bullet$ {\it Remarkable Sequential Mergers}: A particularly striking result from the ASTRID simulation is the identification of two high-detection-probability CW events arising from two subsequent mergers within the same cluster core, predicted at frequencies of 2 and 10 nanohertz. This triple merger scenario, found within the simulation, represents a novel and potentially highly observable event for PTAs.

$\bullet$ {\it High-Detection Probability Sources}: Our analysis indicates that at the current 16.8 yrs of PTA  observations only a small fraction of simulated mergers have a high detection probability. \reviseagain{Only six sources exhibiting a DP larger than 0.1 within the 100 independent realizations, and they are from six different realizations.} These high-DP events, including the sequential mergers, are typically associated with the highest-mass merging systems. 
\reviseagain{This is not only consistent with the predictions made based on Illustris 
\citep{Becsy2022}, but also aligned with the non-detection of the CW sources in the NANOGrav 15yr Data Set \citep{Agazie2023_NANOgrav_SSserch}.
This low detection probability implies the necessity to include more pulsars and to improve the pulsar noise models \citep{Kelley2018_SS, Goncharov2021}. 
}

$\bullet$ {\it Redshift Distribution}: The redshift distribution of all foreground sources is quite spread out, with a peak at $z\sim 0.2$, and a minor peak at $z\sim1.2$ at lower significance as it is due to only two high-DP systems. When weighted by gravitational wave energy, a single peak emerges at $z \sim 0.2$.

$\bullet$ {\it SMBH Merger Properties}: Most foreground sources are major mergers with mass ratios of $q > 0.1$ and a median $q$ of around 0.5. The total MBH mass of these mergers is typically high, with almost no events with chirp masses below $\sim 10^9 \mathrm{M}_{\odot}$ detectable.

$\bullet$ {\it Link to (Dual) AGN}: We find a strong association between foreground events and AGN activity. The majority (87.4\%) of foreground events are associated with dual AGN, where both black holes in the merging pair have high bolometric luminosity ($L_{\mathrm{bol}} >10^{43}$ erg/s). However, only a small fraction (0.7\%) of dual AGNs with $M_{\mathrm{BH}}>10^{8}\,\mathrm{M}_{\odot}$ at $z = 0.3$ evolves into foreground sources later.

$\bullet$ {\it Host Galaxy Properties}: The host galaxies of the high-DP foreground events are generally massive, with stellar masses typically exceeding $10^{12}\,\mathrm{M}_{\odot}$. \revise{The six host galaxies all have a star formation rate over $5\,\mathrm{M}_{\odot}/$yr within 20~kpc from the galaxy's center.
The galaxy hosting the triple merger goes through very significant star formation activity, whose SFR is over $100\ \mathrm{M}_{\odot}/{\rm yr}$.}


Given our results, what guidance can we offer in searches for the  EM counterparts of individual PTA sources? We have seen that in \astrid\ all detectable sources are in massive galaxies, the most prominent being the host of two events lying at the center of a galaxy cluster of mass $10^{15}$ \Msun at
$z=0.25$. Such objects have highly biased clustering (e.g., \citealt{desjacques18}, \citealt{delucia07}), leading to
an expected anisotropy in the GWB. Whether they can be localized in the future depends on the signal-to-noise ratio (SNR). As an example,  \cite{truant24} modeled the detectability of individual events with idealized MeerKAT and SKA
experiments, finding 
that binaries
detected at $5 < {\rm SNR} < 10$ have a median sky-localization area of
$\Delta \Omega \sim 200$ deg$^2$ (falling to $\sim 10\, {\rm deg}^2$
for SNR $>$ 15).
As these authors stated, this error ellipse of $\sim 1-10$ deg radius 
makes localization challenging. On the other hand, the number of galaxy clusters with mass $\geq 10^{15}$\Msun and $z < 0.25$ is
less than 100 (\citealt{abell1958,reiprich02,bohringer17}). This leads to a mean angular  separation of  $>20$ deg and consequently a low chance of confusion.

In \astrid\ the triple merger occurs in a BCG. This and several other detectable GW events take place in systems that are not “dry” mergers; these galaxies exhibit substantial star formation rates, on the order of hundreds of solar masses per year. Star-forming populations of BCGs have been observed at these moderate redshifts, for example, in the KIDS survey \cite{Castignani22}.
While the presence of (dual) AGN appears to be a necessary condition for a merger to become a foreground GW event, we find that fewer than 1\% of dual AGN systems evolve into high-DP events. 
We predict that a source is typically located in a galaxy at the center of a cluster that has undergone recent or still ongoing mergers—either through the merger of at least two large galaxies or cannibalism of a smaller galaxy by a larger elliptical. A key detail is that all systems we have found retain enough residual gas to fuel a high total star formation rate. This is another factor that could enhance the detectability of the associated CW signal.

Our study validates the theoretical framework by showing similarities to other simulation based works and observational constraints.
These results reinforce the idea that single source detections, in addition to the GWB, are likely and important for probing the population and evolution of SMBHBs. The possible detection of CWs in current/upcoming data, combined with the growing understanding of GWB, further refines our understanding of the parameter space. This research underscores the importance of considering foreground sources when interpreting PTA data, and helps to guide future multi-messenger searches for the EM counterparts of these events.
To better constrain theoretical predictions, future works should focus on improving the modeling of binary evolution, the inclusion of more realistic noise models, and the exploration of the degeneracies between model parameters.

\section*{Acknowledgements}
TDM acknowledges funding from NASA ATP 80NSSC20K0519, NSF PHY-2020295, NASA ATP NNX17AK56G, and NASA ATP 80NSSC18K101, NASA Theory grant 80NSSC22K072.
NC acknowledges support from the Schmidt Futures Fund. 
YN acknowledges support from the ITC Postdoctoral Fellowship.
SB acknowledges funding from NASA ATP 80NSSC22K1897.
\astrid~was run on the Frontera facility at the Texas Advanced Computing Center.
 
\section*{Data Availability}
\revise{The code to reproduce the simulation is available at \url{https://github.com/MP-Gadget/MP-Gadget}, and continues to be developed.
The \astrid\ snapshots are available at \url{https://astrid-portal.psc.edu/}.
The merger catalog is available upon request.}

\appendix


\section{Detection Probability}
\label{app:DP}
We calculate the detection probability (DP) following the prescription provided in \citet{Rosado2015}.
We include 68 pulsars with the 16.03 yr duration, consistent with \citet{Agazie2023_Nanograv_GWB}. The position of pulsars and CW sources are randomly generated, and uniformly distributed across the sky. 
For the noise spectral density of pulsars, we include both white noise and red noise: 
\begin{equation}
    S = 2 \Delta t\, \sigma^{2}_{\mathrm{WN}} + \frac{h^{2}_{\mathrm{rest}}}{12 \pi^{2} f^{3}},
\end{equation}
where the term $2\Delta t\, \sigma_{\mathrm{WN}}^{2}$ is the contribution from the white noise. $\Delta t$ is the observing time cadence, which is set to be $\Delta t = 0.05$~yr in this work. 
We use the white noise root-mean-square $\sigma_{\mathrm{WN}}=3\ \mu s$, which is consistent with the model $d$ used in \citet{Kelley2018_SS}. 
For the red noise term, $h_{\mathrm{rest}}$ is the sum of characteristic strain from all the other single sources as well as the GWB at this frequency:
\begin{equation}
    h_{\mathrm{rest,j}}^{2} = h_{\mathrm{c,BG}}^{2} + \sum_{i\neq j} h_{\mathrm{c,SS}}.
 \end{equation}
 With the noise model, we then calculated the DP for each single source ($\mathrm{DP}_{\mathrm{SS,j}}$). The probability of detecting any of a single source is
 \begin{equation}
     \mathrm{DP}_{\mathrm{SS}} = 1-\prod_{j}\left[1-\mathrm{DP}_{\mathrm{SS,j}}\right].
 \end{equation}
As pointed out in \citet{Mingarelli2017}, the relative position between single sources and the pulsars on the sky has a significant influence on the resultant expected number of the detected sources.
To account for the effect of observed sky-position, we generated 500 sets of parameters related to observation (rather than the GW intrinsic properties). This is the so-called `sky realization' in \citet{Gardiner2024_PTACW}. Each set includes a randomly assigned pulsar position, single source position, inclination, polarization, and the GW phase. The DP for a CW source that occurred in a specific strain realization is the DP value averaged over 500 observation sets.


\bibliography{reference}{}

\begin{thebibliography}{}
\expandafter\ifx\csname natexlab\endcsname\relax\def\natexlab#1{#1}\fi
\providecommand{\url}[1]{\href{#1}{#1}}
\providecommand{\dodoi}[1]{doi:~\href{http://doi.org/#1}{\nolinkurl{#1}}}
\providecommand{\doeprint}[1]{\href{http://ascl.net/#1}{\nolinkurl{http://ascl.net/#1}}}
\providecommand{\doarXiv}[1]{\href{https://arxiv.org/abs/#1}{\nolinkurl{https://arxiv.org/abs/#1}}}

\bibitem[{{Abell}(1958)}]{abell1958}
{Abell}, G.~O. 1958, \apjs, 3, 211, \dodoi{10.1086/190036}

\bibitem[{{Agazie} {et~al.}(2023{\natexlab{a}}){Agazie}, {Anumarlapudi}, {Archibald}, {Arzoumanian}, {Baker}, {B{\'e}csy}, {Blecha}, {Brazier}, {Brook}, {Burke-Spolaor}, {Burnette}, {Case}, {Charisi}, {Chatterjee}, {Chatziioannou}, {Cheeseboro}, {Chen}, {Cohen}, {Cordes}, {Cornish}, {Crawford}, {Cromartie}, {Crowter}, {Cutler}, {Decesar}, {Degan}, {Demorest}, {Deng}, {Dolch}, {Drachler}, {Ellis}, {Ferrara}, {Fiore}, {Fonseca}, {Freedman}, {Garver-Daniels}, {Gentile}, {Gersbach}, {Glaser}, {Good}, {G{\"u}ltekin}, {Hazboun}, {Hourihane}, {Islo}, {Jennings}, {Johnson}, {Jones}, {Kaiser}, {Kaplan}, {Kelley}, {Kerr}, {Key}, {Klein}, {Laal}, {Lam}, {Lamb}, {Lazio}, {Lewandowska}, {Littenberg}, {Liu}, {Lommen}, {Lorimer}, {Luo}, {Lynch}, {Ma}, {Madison}, {Mattson}, {McEwen}, {McKee}, {McLaughlin}, {McMann}, {Meyers}, {Meyers}, {Mingarelli}, {Mitridate}, {Natarajan}, {Ng}, {Nice}, {Ocker}, {Olum}, {Pennucci}, {Perera}, {Petrov}, {Pol}, {Radovan}, {Ransom}, {Ray}, {Romano}, {Sardesai}, {Schmiedekamp}, {Schmiedekamp},
  {Schmitz}, {Schult}, {Shapiro-Albert}, {Siemens}, {Simon}, {Siwek}, {Stairs}, {Stinebring}, {Stovall}, {Sun}, {Susobhanan}, {Swiggum}, {Taylor}, {Taylor}, {Turner}, {Unal}, {Vallisneri}, {van Haasteren}, {Vigeland}, {Wahl}, {Wang}, {Witt}, {Young}, \& {Nanograv Collaboration}}]{Agazie2023_Nanograv_GWB}
{Agazie}, G., {Anumarlapudi}, A., {Archibald}, A.~M., {et~al.} 2023{\natexlab{a}}, \apjl, 951, L8, \dodoi{10.3847/2041-8213/acdac6}

\bibitem[{{Agazie} {et~al.}(2023{\natexlab{b}}){Agazie}, {Anumarlapudi}, {Archibald}, {Arzoumanian}, {Baker}, {B{\'e}csy}, {Blecha}, {Brazier}, {Brook}, {Burke-Spolaor}, {Case}, {Casey-Clyde}, {Charisi}, {Chatterjee}, {Cohen}, {Cordes}, {Cornish}, {Crawford}, {Cromartie}, {Crowter}, {Decesar}, {Demorest}, {Digman}, {Dolch}, {Drachler}, {Ferrara}, {Fiore}, {Fonseca}, {Freedman}, {Garver-Daniels}, {Gentile}, {Glaser}, {Good}, {G{\"u}ltekin}, {Hazboun}, {Hourihane}, {Jennings}, {Johnson}, {Jones}, {Kaiser}, {Kaplan}, {Kelley}, {Kerr}, {Key}, {Laal}, {Lam}, {Lamb}, {Lazio}, {Lewandowska}, {Liu}, {Lorimer}, {Luo}, {Lynch}, {Ma}, {Madison}, {McEwen}, {McKee}, {McLaughlin}, {McMann}, {Meyers}, {Meyers}, {Mingarelli}, {Mitridate}, {Ng}, {Nice}, {Ocker}, {Olum}, {Pennucci}, {Perera}, {Petrov}, {Pol}, {Radovan}, {Ransom}, {Ray}, {Romano}, {Sardesai}, {Schmiedekamp}, {Schmiedekamp}, {Schmitz}, {Shapiro-Albert}, {Siemens}, {Simon}, {Siwek}, {Stairs}, {Stinebring}, {Stovall}, {Susobhanan}, {Swiggum}, {Taylor}, {Taylor},
  {Turner}, {Unal}, {Vallisneri}, {van Haasteren}, {Vigeland}, {Wahl}, {Witt}, {Young}, \& {Nanograv Collaboration}}]{Agazie2023_NANOgrav_SSserch}
---. 2023{\natexlab{b}}, \apjl, 951, L50, \dodoi{10.3847/2041-8213/ace18a}

\bibitem[{{Agazie} {et~al.}(2023{\natexlab{c}}){Agazie}, {Anumarlapudi}, {Archibald}, {Arzoumanian}, {Baker}, {B{\'e}csy}, {Blecha}, {Brazier}, {Brook}, {Burke-Spolaor}, {Casey-Clyde}, {Charisi}, {Chatterjee}, {Cohen}, {Cordes}, {Cornish}, {Crawford}, {Cromartie}, {Crowter}, {DeCesar}, {Demorest}, {Dolch}, {Drachler}, {Ferrara}, {Fiore}, {Fonseca}, {Freedman}, {Gardiner}, {Garver-Daniels}, {Gentile}, {Glaser}, {Good}, {G{\"u}ltekin}, {Hazboun}, {Jennings}, {Johnson}, {Jones}, {Kaiser}, {Kaplan}, {Kelley}, {Kerr}, {Key}, {Laal}, {Lam}, {Lamb}, {Lazio}, {Lewandowska}, {Liu}, {Lorimer}, {Luo}, {Lynch}, {Ma}, {Madison}, {McEwen}, {McKee}, {McLaughlin}, {McMann}, {Meyers}, {Mingarelli}, {Mitridate}, {Ng}, {Nice}, {Ocker}, {Olum}, {Pennucci}, {Perera}, {Pol}, {Radovan}, {Ransom}, {Ray}, {Romano}, {Sardesai}, {Schmiedekamp}, {Schmiedekamp}, {Schmitz}, {Schult}, {Shapiro-Albert}, {Siemens}, {Simon}, {Siwek}, {Stairs}, {Stinebring}, {Stovall}, {Susobhanan}, {Swiggum}, {Taylor}, {Turner}, {Unal}, {Vallisneri},
  {Vigeland}, {Wahl}, {Witt}, \& {Young}}]{Agazie2023_anisotropy_search}
---. 2023{\natexlab{c}}, \apjl, 956, L3, \dodoi{10.3847/2041-8213/acf4fd}

\bibitem[{{Agazie} {et~al.}(2023{\natexlab{d}}){Agazie}, {Anumarlapudi}, {Archibald}, {Baker}, {B{\'e}csy}, {Blecha}, {Bonilla}, {Brazier}, {Brook}, {Burke-Spolaor}, {Burnette}, {Case}, {Casey-Clyde}, {Charisi}, {Chatterjee}, {Chatziioannou}, {Cheeseboro}, {Chen}, {Cohen}, {Cordes}, {Cornish}, {Crawford}, {Cromartie}, {Crowter}, {Cutler}, {D'Orazio}, {Decesar}, {Degan}, {Demorest}, {Deng}, {Dolch}, {Drachler}, {Ferrara}, {Fiore}, {Fonseca}, {Freedman}, {Gardiner}, {Garver-Daniels}, {Gentile}, {Gersbach}, {Glaser}, {Good}, {G{\"u}ltekin}, {Hazboun}, {Hourihane}, {Islo}, {Jennings}, {Johnson}, {Jones}, {Kaiser}, {Kaplan}, {Kelley}, {Kerr}, {Key}, {Laal}, {Lam}, {Lamb}, {Lazio}, {Lewandowska}, {Littenberg}, {Liu}, {Luo}, {Lynch}, {Ma}, {Madison}, {McEwen}, {McKee}, {McLaughlin}, {McMann}, {Meyers}, {Meyers}, {Mingarelli}, {Mitridate}, {Natarajan}, {Ng}, {Nice}, {Ocker}, {Olum}, {Pennucci}, {Perera}, {Petrov}, {Pol}, {Radovan}, {Ransom}, {Ray}, {Romano}, {Runnoe}, {Sardesai}, {Schmiedekamp}, {Schmiedekamp},
  {Schmitz}, {Schult}, {Shapiro-Albert}, {Siemens}, {Simon}, {Siwek}, {Stairs}, {Stinebring}, {Stovall}, {Sun}, {Susobhanan}, {Swiggum}, {Taylor}, {Taylor}, {Turner}, {Unal}, {Vallisneri}, {Vigeland}, {Wachter}, {Wahl}, {Wang}, {Witt}, {Wright}, {Young}, \& {Nanograv Collaboration}}]{Agazie2023_GWB_smbh_constrain}
---. 2023{\natexlab{d}}, \apjl, 952, L37, \dodoi{10.3847/2041-8213/ace18b}

\bibitem[{{Agazie} {et~al.}(2024){Agazie}, {Antoniadis}, {Anumarlapudi}, {Archibald}, {Arumugam}, {Arumugam}, {Arzoumanian}, {Askew}, {Babak}, {Bagchi}, {Bailes}, {Bak Nielsen}, {Baker}, {Bassa}, {Bathula}, {B{\'e}csy}, {Berthereau}, {Bhat}, {Blecha}, {Bonetti}, {Bortolas}, {Brazier}, {Brook}, {Burgay}, {Burke-Spolaor}, {Burnette}, {Caballero}, {Cameron}, {Case}, {Chalumeau}, {Champion}, {Chanlaridis}, {Charisi}, {Chatterjee}, {Chatziioannou}, {Cheeseboro}, {Chen}, {Chen}, {Cognard}, {Cohen}, {Coles}, {Cordes}, {Cornish}, {Crawford}, {Cromartie}, {Crowter}, {Cury{\l}o}, {Cutler}, {Dai}, {Dandapat}, {Deb}, {DeCesar}, {DeGan}, {Demorest}, {Deng}, {Desai}, {Desvignes}, {Dey}, {Dhanda-Batra}, {Di Marco}, {Dolch}, {Drachler}, {Dwivedi}, {Ellis}, {Falxa}, {Feng}, {Ferdman}, {Ferrara}, {Fiore}, {Fonseca}, {Franchini}, {Freedman}, {Gair}, {Garver-Daniels}, {Gentile}, {Gersbach}, {Glaser}, {Good}, {Goncharov}, {Gopakumar}, {Graikou}, {Griessmeier}, {Guillemot}, {G{\"u}ltekin}, {Guo}, {Gupta}, {Grunthal}, {Hazboun},
  {Hisano}, {Hobbs}, {Hourihane}, {Hu}, {Iraci}, {Islo}, {Izquierdo-Villalba}, {Jang}, {Jawor}, {Janssen}, {Jennings}, {Jessner}, {Johnson}, {Jones}, {Joshi}, {Kaiser}, {Kaplan}, {Kapur}, {Kareem}, {Karuppusamy}, {Keane}, {Keith}, {Kelley}, {Kerr}, {Key}, {Kharbanda}, {Kikunaga}, {Klein}, {Kolhe}, {Kramer}, {Krishnakumar}, {Kulkarni}, {Laal}, {Lackeos}, {Lam}, {Lamb}, {Larsen}, {Lazio}, {Lee}, {Levin}, {Lewandowska}, {Littenberg}, {Liu}, {Liu}, {Liu}, {Lommen}, {Lorimer}, {Lower}, {Luo}, {Luo}, {Lynch}, {Lyne}, {Ma}, {Maan}, {Madison}, {Main}, {Manchester}, {Mandow}, {Mattson}, {McEwen}, {McKee}, {McLaughlin}, {McMann}, {Meyers}, {Meyers}, {Mickaliger}, {Miles}, {Mingarelli}, {Mitridate}, {Natarajan}, {Nathan}, {Ng}, {Nice}, {Ni{\c{t}}u}, {Nobleson}, {Ocker}, {Olum}, {Os{\l}owski}, {Paladi}, {Parthasarathy}, {Pennucci}, {Perera}, {Perrodin}, {Petiteau}, {Petrov}, {Pol}, {Porayko}, {Possenti}, {Prabu}, {Quelquejay Leclere}, {Radovan}, {Rana}, {Ransom}, {Ray}, {Reardon}, {Rogers}, {Romano}, {Russell},
  {Samajdar}, {Sanidas}, {Sardesai}, {Schmiedekamp}, {Schmiedekamp}, {Schmitz}, {Schult}, {Sesana}, {Shaifullah}, {Shannon}, {Shapiro-Albert}, {Siemens}, {Simon}, \& {Singha}}]{Agazie2024_PTAcompare}
{Agazie}, G., {Antoniadis}, J., {Anumarlapudi}, A., {et~al.} 2024, \apj, 966, 105, \dodoi{10.3847/1538-4357/ad36be}

\bibitem[{{B{\'e}csy} {et~al.}(2022){B{\'e}csy}, {Cornish}, \& {Kelley}}]{Becsy2022}
{B{\'e}csy}, B., {Cornish}, N.~J., \& {Kelley}, L.~Z. 2022, \apj, 941, 119, \dodoi{10.3847/1538-4357/aca1b2}

\bibitem[{{Bird} {et~al.}(2022){Bird}, {Ni}, {Di Matteo}, {Croft}, {Feng}, \& {Chen}}]{Bird2022_astrid}
{Bird}, S., {Ni}, Y., {Di Matteo}, T., {et~al.} 2022, \mnras, 512, 3703, \dodoi{10.1093/mnras/stac648}

\bibitem[{{B{\"o}hringer} {et~al.}(2017){B{\"o}hringer}, {Chon}, \& {Fukugita}}]{bohringer17}
{B{\"o}hringer}, H., {Chon}, G., \& {Fukugita}, M. 2017, \aap, 608, A65, \dodoi{10.1051/0004-6361/201731205}

\bibitem[{{Castignani} {et~al.}(2022){Castignani}, {Radovich}, {Combes}, {Salom{\'e}}, {Maturi}, {Moscardini}, {Bardelli}, {Giocoli}, {Lesci}, {Marulli}, {Puddu}, \& {Sereno}}]{Castignani22}
{Castignani}, G., {Radovich}, M., {Combes}, F., {et~al.} 2022, \aap, 667, A52, \dodoi{10.1051/0004-6361/202243689}

\bibitem[{{Chen} {et~al.}(2025){Chen}, {DiMatteo}, \& {Zhou}}]{PTA_astrid_GWB}
{Chen}, N., {DiMatteo}, T., \& {Zhou}, Y. 2025, \apj.
\newblock \doarXiv{1302.4485}

\bibitem[{{Chen} {et~al.}(2022){Chen}, {Ni}, {Tremmel}, {Di Matteo}, {Bird}, {DeGraf}, \& {Feng}}]{Chen2022_DFmodel}
{Chen}, N., {Ni}, Y., {Tremmel}, M., {et~al.} 2022, \mnras, 510, 531, \dodoi{10.1093/mnras/stab3411}

\bibitem[{{Chen} {et~al.}(2023){Chen}, {Di Matteo}, {Ni}, {Tremmel}, {DeGraf}, {Shen}, {Holgado}, {Bird}, {Croft}, \& {Feng}}]{Chen2023_dualAGN}
{Chen}, N., {Di Matteo}, T., {Ni}, Y., {et~al.} 2023, \mnras, 522, 1895, \dodoi{10.1093/mnras/stad834}

\bibitem[{{De Lucia} \& {Blaizot}(2007)}]{delucia07}
{De Lucia}, G., \& {Blaizot}, J. 2007, \mnras, 375, 2, \dodoi{10.1111/j.1365-2966.2006.11287.x}

\bibitem[{{Desjacques} {et~al.}(2018){Desjacques}, {Jeong}, \& {Schmidt}}]{desjacques18}
{Desjacques}, V., {Jeong}, D., \& {Schmidt}, F. 2018, \physrep, 733, 1, \dodoi{10.1016/j.physrep.2017.12.002}

\bibitem[{{EPTA Collaboration} {et~al.}(2023){EPTA Collaboration}, {InPTA Collaboration}, {Antoniadis}, {Arumugam}, {Arumugam}, {Babak}, {Bagchi}, {Bak Nielsen}, {Bassa}, {Bathula}, {Berthereau}, {Bonetti}, {Bortolas}, {Brook}, {Burgay}, {Caballero}, {Chalumeau}, {Champion}, {Chanlaridis}, {Chen}, {Cognard}, {Dandapat}, {Deb}, {Desai}, {Desvignes}, {Dhanda-Batra}, {Dwivedi}, {Falxa}, {Ferdman}, {Franchini}, {Gair}, {Goncharov}, {Gopakumar}, {Graikou}, {Grie{\ss}meier}, {Guillemot}, {Guo}, {Gupta}, {Hisano}, {Hu}, {Iraci}, {Izquierdo-Villalba}, {Jang}, {Jawor}, {Janssen}, {Jessner}, {Joshi}, {Kareem}, {Karuppusamy}, {Keane}, {Keith}, {Kharbanda}, {Kikunaga}, {Kolhe}, {Kramer}, {Krishnakumar}, {Lackeos}, {Lee}, {Liu}, {Liu}, {Lyne}, {McKee}, {Maan}, {Main}, {Mickaliger}, {Ni{\c{t}}u}, {Nobleson}, {Paladi}, {Parthasarathy}, {Perera}, {Perrodin}, {Petiteau}, {Porayko}, {Possenti}, {Prabu}, {Quelquejay Leclere}, {Rana}, {Samajdar}, {Sanidas}, {Sesana}, {Shaifullah}, {Singha}, {Speri}, {Spiewak}, {Srivastava},
  {Stappers}, {Surnis}, {Susarla}, {Susobhanan}, {Takahashi}, {Tarafdar}, {Theureau}, {Tiburzi}, {van der Wateren}, {Vecchio}, {Venkatraman Krishnan}, {Verbiest}, {Wang}, {Wang}, \& {Wu}}]{EPTACollaboration2023}
{EPTA Collaboration}, {InPTA Collaboration}, {Antoniadis}, J., {et~al.} 2023, \aap, 678, A50, \dodoi{10.1051/0004-6361/202346844}

\bibitem[{{Fastidio} {et~al.}(2024){Fastidio}, {Gualandris}, {Sesana}, {Bortolas}, \& {Dehnen}}]{Fastidio2024}
{Fastidio}, F., {Gualandris}, A., {Sesana}, A., {Bortolas}, E., \& {Dehnen}, W. 2024, \mnras, 532, 295, \dodoi{10.1093/mnras/stae1411}

\bibitem[{{Franchini} {et~al.}(2024){Franchini}, {Prato}, {Longarini}, \& {Sesana}}]{Franchini2024}
{Franchini}, A., {Prato}, A., {Longarini}, C., \& {Sesana}, A. 2024, \aap, 688, A174, \dodoi{10.1051/0004-6361/202449402}

\bibitem[{{Gardiner} {et~al.}(2024){Gardiner}, {Kelley}, {Lemke}, \& {Mitridate}}]{Gardiner2024_PTACW}
{Gardiner}, E.~C., {Kelley}, L.~Z., {Lemke}, A.-M., \& {Mitridate}, A. 2024, \apj, 965, 164, \dodoi{10.3847/1538-4357/ad2be8}

\bibitem[{{Genel} {et~al.}(2014){Genel}, {Vogelsberger}, {Springel}, {Sijacki}, {Nelson}, {Snyder}, {Rodriguez-Gomez}, {Torrey}, \& {Hernquist}}]{Genel2014_illustris}
{Genel}, S., {Vogelsberger}, M., {Springel}, V., {et~al.} 2014, \mnras, 445, 175, \dodoi{10.1093/mnras/stu1654}

\bibitem[{{Genina} {et~al.}(2024){Genina}, {Springel}, \& {Rantala}}]{Genina2024}
{Genina}, A., {Springel}, V., \& {Rantala}, A. 2024, \mnras, 534, 957, \dodoi{10.1093/mnras/stae2144}

\bibitem[{{Goncharov} {et~al.}(2021){Goncharov}, {Reardon}, {Shannon}, {Zhu}, {Thrane}, {Bailes}, {Bhat}, {Dai}, {Hobbs}, {Kerr}, {Manchester}, {Os{\l}owski}, {Parthasarathy}, {Russell}, {Spiewak}, {Thyagarajan}, \& {Wang}}]{Goncharov2021}
{Goncharov}, B., {Reardon}, D.~J., {Shannon}, R.~M., {et~al.} 2021, \mnras, 502, 478, \dodoi{10.1093/mnras/staa3411}

\bibitem[{{Hsu} {et~al.}(2022){Hsu}, {Lin}, {Huang}, {Nelson}, {Rodriguez-Gomez}, {Lai}, {Greene}, {Leauthaud}, {Arag{\'o}n-Salamanca}, {Bundy}, {Emsellem}, {Merrifield}, {More}, {Okabe}, {Rong}, {Brownstein}, {Lane}, {Pan}, \& {Schneider}}]{hsu22}
{Hsu}, Y.-H., {Lin}, Y.-T., {Huang}, S., {et~al.} 2022, \apj, 933, 61, \dodoi{10.3847/1538-4357/ac6d66}

\bibitem[{{Izquierdo-Villalba} {et~al.}(2022){Izquierdo-Villalba}, {Sesana}, {Bonoli}, \& {Colpi}}]{Izquierdo-Villalba2022}
{Izquierdo-Villalba}, D., {Sesana}, A., {Bonoli}, S., \& {Colpi}, M. 2022, \mnras, 509, 3488, \dodoi{10.1093/mnras/stab3239}

\bibitem[{{Kelley} {et~al.}(2017){Kelley}, {Blecha}, \& {Hernquist}}]{Kelley2017_ills_env}
{Kelley}, L.~Z., {Blecha}, L., \& {Hernquist}, L. 2017, \mnras, 464, 3131, \dodoi{10.1093/mnras/stw2452}

\bibitem[{{Kelley} {et~al.}(2018){Kelley}, {Blecha}, {Hernquist}, {Sesana}, \& {Taylor}}]{Kelley2018_SS}
{Kelley}, L.~Z., {Blecha}, L., {Hernquist}, L., {Sesana}, A., \& {Taylor}, S.~R. 2018, \mnras, 477, 964, \dodoi{10.1093/mnras/sty689}

\bibitem[{{Mingarelli} {et~al.}(2017){Mingarelli}, {Lazio}, {Sesana}, {Greene}, {Ellis}, {Ma}, {Croft}, {Burke-Spolaor}, \& {Taylor}}]{Mingarelli2017}
{Mingarelli}, C. M.~F., {Lazio}, T. J.~W., {Sesana}, A., {et~al.} 2017, Nature Astronomy, 1, 886, \dodoi{10.1038/s41550-017-0299-6}

\bibitem[{{Ni} {et~al.}(2024){Ni}, {Chen}, {Zhou}, {Park}, {Yang}, {DiMatteo}, {Bird}, \& {Croft}}]{Ni2024}
{Ni}, Y., {Chen}, N., {Zhou}, Y., {et~al.} 2024, arXiv e-prints, arXiv:2409.10666, \dodoi{10.48550/arXiv.2409.10666}

\bibitem[{{Ni} {et~al.}(2022){Ni}, {Di Matteo}, {Bird}, {Croft}, {Feng}, {Chen}, {Tremmel}, {DeGraf}, \& {Li}}]{Ni2022_astrid}
{Ni}, Y., {Di Matteo}, T., {Bird}, S., {et~al.} 2022, \mnras, 513, 670, \dodoi{10.1093/mnras/stac351}

\bibitem[{{Ravi} {et~al.}(2012){Ravi}, {Wyithe}, {Hobbs}, {Shannon}, {Manchester}, {Yardley}, \& {Keith}}]{Ravi2012}
{Ravi}, V., {Wyithe}, J.~S.~B., {Hobbs}, G., {et~al.} 2012, \apj, 761, 84, \dodoi{10.1088/0004-637X/761/2/84}

\bibitem[{{Reardon} {et~al.}(2023){Reardon}, {Zic}, {Shannon}, {Hobbs}, {Bailes}, {Di Marco}, {Kapur}, {Rogers}, {Thrane}, {Askew}, {Bhat}, {Cameron}, {Cury{\l}o}, {Coles}, {Dai}, {Goncharov}, {Kerr}, {Kulkarni}, {Levin}, {Lower}, {Manchester}, {Mandow}, {Miles}, {Nathan}, {Os{\l}owski}, {Russell}, {Spiewak}, {Zhang}, \& {Zhu}}]{Reardon2023_PPTA}
{Reardon}, D.~J., {Zic}, A., {Shannon}, R.~M., {et~al.} 2023, \apjl, 951, L6, \dodoi{10.3847/2041-8213/acdd02}

\bibitem[{{Reines} \& {Volonteri}(2015)}]{Reines2015}
{Reines}, A.~E., \& {Volonteri}, M. 2015, \apj, 813, 82, \dodoi{10.1088/0004-637X/813/2/82}

\bibitem[{{Reiprich} \& {B{\"o}hringer}(2002)}]{reiprich02}
{Reiprich}, T.~H., \& {B{\"o}hringer}, H. 2002, \apj, 567, 716, \dodoi{10.1086/338753}

\bibitem[{{Rodriguez-Gomez} {et~al.}(2015){Rodriguez-Gomez}, {Genel}, {Vogelsberger}, {Sijacki}, {Pillepich}, {Sales}, {Torrey}, {Snyder}, {Nelson}, {Springel}, {Ma}, \& {Hernquist}}]{Rodriguez-Gomez2015_illu_mergingrate}
{Rodriguez-Gomez}, V., {Genel}, S., {Vogelsberger}, M., {et~al.} 2015, \mnras, 449, 49, \dodoi{10.1093/mnras/stv264}

\bibitem[{{Roedig} \& {Sesana}(2012)}]{Roedig2012}
{Roedig}, C., \& {Sesana}, A. 2012, in Journal of Physics Conference Series, Vol. 363, Journal of Physics Conference Series (IOP), 012035, \dodoi{10.1088/1742-6596/363/1/012035}

\bibitem[{{Rosado} {et~al.}(2015){Rosado}, {Sesana}, \& {Gair}}]{Rosado2015}
{Rosado}, P.~A., {Sesana}, A., \& {Gair}, J. 2015, \mnras, 451, 2417, \dodoi{10.1093/mnras/stv1098}

\bibitem[{{Saeedzadeh} {et~al.}(2024){Saeedzadeh}, {Mukherjee}, {Babul}, {Tremmel}, \& {Quinn}}]{Saeedzadeh2024}
{Saeedzadeh}, V., {Mukherjee}, S., {Babul}, A., {Tremmel}, M., \& {Quinn}, T.~R. 2024, \mnras, 529, 4295, \dodoi{10.1093/mnras/stae513}

\bibitem[{{Sah} {et~al.}(2024){Sah}, {Mukherjee}, {Saeedzadeh}, {Babul}, {Tremmel}, \& {Quinn}}]{Sah2024}
{Sah}, M.~R., {Mukherjee}, S., {Saeedzadeh}, V., {et~al.} 2024, \mnras, 533, 1568, \dodoi{10.1093/mnras/stae1930}

\bibitem[{{Sato-Polito} \& {Zaldarriaga}(2024)}]{Sato-Polito2024_dist}
{Sato-Polito}, G., \& {Zaldarriaga}, M. 2024, arXiv e-prints, arXiv:2406.17010, \dodoi{10.48550/arXiv.2406.17010}

\bibitem[{{Sato-Polito} {et~al.}(2024){Sato-Polito}, {Zaldarriaga}, \& {Quataert}}]{Sato-Polito2024_diff}
{Sato-Polito}, G., {Zaldarriaga}, M., \& {Quataert}, E. 2024, \prd, 110, 063020, \dodoi{10.1103/PhysRevD.110.063020}

\bibitem[{{Sesana}(2010)}]{Sesana2010}
{Sesana}, A. 2010, \apj, 719, 851, \dodoi{10.1088/0004-637X/719/1/851}

\bibitem[{{Sesana} {et~al.}(2008){Sesana}, {Vecchio}, \& {Colacino}}]{Sesana2008}
{Sesana}, A., {Vecchio}, A., \& {Colacino}, C.~N. 2008, \mnras, 390, 192, \dodoi{10.1111/j.1365-2966.2008.13682.x}

\bibitem[{{Shakura} \& {Sunyaev}(1973)}]{Shakura1973_BH}
{Shakura}, N.~I., \& {Sunyaev}, R.~A. 1973, \aap, 24, 337

\bibitem[{{Skrutskie} {et~al.}(2006){Skrutskie}, {Cutri}, {Stiening}, {Weinberg}, {Schneider}, {Carpenter}, {Beichman}, {Capps}, {Chester}, {Elias}, {Huchra}, {Liebert}, {Lonsdale}, {Monet}, {Price}, {Seitzer}, {Jarrett}, {Kirkpatrick}, {Gizis}, {Howard}, {Evans}, {Fowler}, {Fullmer}, {Hurt}, {Light}, {Kopan}, {Marsh}, {McCallon}, {Tam}, {Van Dyk}, \& {Wheelock}}]{Skrutskie2006_2mass}
{Skrutskie}, M.~F., {Cutri}, R.~M., {Stiening}, R., {et~al.} 2006, \aj, 131, 1163, \dodoi{10.1086/498708}

\bibitem[{{Tremmel} {et~al.}(2015){Tremmel}, {Governato}, {Volonteri}, \& {Quinn}}]{Tremmel2015_DFmodel}
{Tremmel}, M., {Governato}, F., {Volonteri}, M., \& {Quinn}, T.~R. 2015, \mnras, 451, 1868, \dodoi{10.1093/mnras/stv1060}

\bibitem[{{Truant} {et~al.}(2024){Truant}, {Izquierdo-Villalba}, {Sesana}, {Mohiuddin Shaifullah}, \& {Bonetti}}]{truant24}
{Truant}, R.~J., {Izquierdo-Villalba}, D., {Sesana}, A., {Mohiuddin Shaifullah}, G., \& {Bonetti}, M. 2024, arXiv e-prints, arXiv:2407.12078, \dodoi{10.48550/arXiv.2407.12078}

\bibitem[{{Xu} {et~al.}(2023){Xu}, {Chen}, {Guo}, {Jiang}, {Wang}, {Xu}, {Xue}, {Nicolas Caballero}, {Yuan}, {Xu}, {Wang}, {Hao}, {Luo}, {Lee}, {Han}, {Jiang}, {Shen}, {Wang}, {Wang}, {Xu}, {Wu}, {Manchester}, {Qian}, {Guan}, {Huang}, {Sun}, \& {Zhu}}]{Xu2023_CPTA}
{Xu}, H., {Chen}, S., {Guo}, Y., {et~al.} 2023, Research in Astronomy and Astrophysics, 23, 075024, \dodoi{10.1088/1674-4527/acdfa5}

\bibitem[{{Zhou} {et~al.}(2025){Zhou}, {Mukherjee}, {Chen}, {Di Matteo}, {Johansson}, {Rantala}, {Partmann}, {Di Carlo}, {Bird}, \& {Ni}}]{Zhou2025}
{Zhou}, Y., {Mukherjee}, D., {Chen}, N., {et~al.} 2025, \apj, 980, 79, \dodoi{10.3847/1538-4357/ada283}

\end{thebibliography}
\bibliographystyle{aasjournal}



\end{document}